\documentclass[showpacs,twocolumn,floatfix,pra]{revtex4}
\usepackage{graphicx}
\usepackage{bm}
\usepackage{psfrag}
\newcommand{\ket}[1]{| #1 \rangle}
\newcommand{\bra}[1]{\langle #1 |}
\newcommand{\rb}[1]{\left( #1 \right)}

\newcommand{\beq}{\begin{eqnarray}}
\newcommand{\eeq}{\end{eqnarray}}

\newcommand{\op}[2]{| #1 \rangle \langle #2 |}
\begin{document}
\title{Entanglement in quantum catastrophes}

\author{Clive Emary 
  \footnote{Present address: 
    Department of Physics, 
    University of California San Diego,
    La Jolla, California 92093-0319
    .}
  }
\affiliation{
              Instituut-Lorentz,
              Universiteit Leiden,
              P. O. Box 9506, 2300 RA Leiden,
              The Netherlands}
\author{Neill Lambert and  Tobias Brandes}
\affiliation{
           School of Physics and Astronomy,
           The University of Manchester,
           P.O. Box 88,
           Manchester
           M60 1QD,
           U.K.}
\date{17th March 2005}
\begin{abstract}
We classify entanglement singularities for various two-mode bosonic
systems in terms of catastrophe theory. Employing an abstract 
phase-space representation, we obtain exact results in limiting cases 
for the entropy in cusp, butterfly, and two-dimensional catastrophes. 
We furthermore use numerical results to extract the scaling of the 
entropy with the non-linearity parameter, and discuss the role of  
mixing entropies in more complex systems.  
\end{abstract}
\pacs{03.67.Mn, 03.65.Ud, 42.50.Fx}
\maketitle

For a large number of quantum critical systems, 
criticality manifests itself as a peak, or indeed a
divergence, in the entanglement of the ground state.  
Systems in which this behaviour has been observed 
include spin-$1/2$ ferromagnetic chains in a  magnetic 
field \cite{Ostetal02}, driven, dissipative  large-$j$ pseudo-spin 
models \cite{SM02}, the Lipkin-Meshkov-Glick Hamiltonian from 
nuclear physics \cite{LMG65,VPM04,DV04,LORV04}, and the Dicke model 
from quantum optics \cite{NLCETB03,NLCETB04, RQJ04}.

The high degree of similarity between the behaviour of these systems 
suggests an underlying universality, and in this paper we
explore this universality in terms of a quantum mechanical 
catastrophe theory.

In its elementary, classical form, catastrophe theory is the study of 
the critical points of potentials, with emphasis on a
qualitative understanding of the properties of the system 
as critical points are born, move about, merge, and disappear as 
control parameters are varied \cite{gil81}.
The best known catastrophe is the {\it cusp}, which describes the
bifurcation of a critical point. 
The relation between entanglement properties and the cusp has been noted
previously for the Dicke model \cite{NLCETB03}, and 
the importance of bifurcations in the appearance of entanglement maxima
has been conjectured as a general rule 
\cite{HMM03,HDMM04}. 
In this paper, we shall explore and expand upon these 
ideas.

Some properties of the quantum cusp have been discussed by Gilmore 
{\it et al.} \cite{gil86} but our focus here is different, and 
the way in which we obtain a quantum model from the 
classical catastrophe differs accordingly. 
The method employed here admits the concept of a macroscopic
or semi-classical limit; thus establishing the 
connexion with  
models of quantum phase transitions (QPT).
The quantum cusp model we construct may be thought of as a minimal model
that exhibits the salient entanglement features observed in these models.  
We study not only the cusp, but two further catastrophes 
--- the butterfly and a two-dimensional example --- the entanglement
properties of which expand upon the types of
behaviour one might expect in more realistic models.

\section{Quantum catastrophe models}

We begin by constructing the quantum catastrophe models and first consider
those derived from catastrophes occurring in a single variable, 
such as the cusp.

We take as our model a system of two interacting bosonic modes.  Let
($x_1$, $p_{x_1}$) and ($x_2$, $p_{x_2}$) be the (abstract) position 
and momentum coordinates representing these modes.
We assume an interaction between these modes such that the interacting 
system is separable in a description in 
terms of two {\it collective} bosonic excitations, 
the coordinates of which we denote $(y_1,p_{y_1})$ and $(y_2,p_{y_2})$.
We construct the Hamiltonian of one of these collective modes $y_1$ 
so that it undergoes  the catastrophe.  
The question that we shall address is then: given the structure 
of the system in terms of the collective modes $\mathbf{y}$, 
what is the entanglement between the original bare modes $\mathbf{x}$?

We write the Hamiltonian of the collective mode 
in which the catastrophe occurs as
\beq
  H_1 = \frac{1}{2m} p_{y_1}^2 + m \omega^2 U_\mathrm{cat} (y_1)
\eeq
with $m$ and $\omega$ the  characteristic mass and 
frequency of the mode.
The potential $U_\mathrm{cat} (y_1)$ is taken from elementary
catastrophe theory, and can be written as a power series 
$U_\mathrm{cat} (y) = \sum_{n=1}^{\infty} A_n y_1^n$.  We rescale the 
coordinate $y_1 \rightarrow y_1 \sqrt{\hbar/m\omega} $, and measure the 
energy in units $\hbar \omega$, such that
\beq
  H_1 &=&-\frac{1}{2} \frac{d^2}{dy_1^2} 
  + \sum_{n=1}^{\infty} \frac{A_n}{\mu^{n/2-1}} y_1^n
  \nonumber \\
  &=& -\frac{1}{2} \frac{d^2}{dy_1^2} + V_\mathrm{cat}(y_1),
  \label{H1}
\eeq
which defines the rescaled catastrophe potential 
$V_\mathrm{cat}(y_1)$.
Here, $\mu \equiv m \omega / \hbar$ is our explicit ``macroscopy'' 
parameter, which 
is meant in the sense that the limit $\mu \to \infty$ can be thought 
of either as the limit in which the system size (and hence mass $m$) becomes
macroscopic, or as the semi-classical limit $\hbar \to 0$.
The limit $\mu\to\infty$ is analogous to the thermodynamic 
limit in the QPT models, and therein lies the correspondence 
between these quantum catastrophes and the QPT work cited in the 
introduction.

The behaviour of the mode described by the $H_1$ 
is largely governed by the fixed points of the classical catastrophe 
potential $V_\mathrm{cat}(y_1)$, and this is especially true in the 
limit  $\mu \to \infty$.  By construction the fixed points of 
$V_\mathrm{cat}(y_1)$, which 
we denote $\tilde{y}$, are of the order $\tilde{y} \sim \sqrt{\mu}$, and 
are thus ``macroscopic''.  Expanding $V_\mathrm{cat}(y_1)$ in Eq. (\ref{H1})
about $\tilde{y}$ and taking the limit  $\mu \to \infty$ we obtain
\beq
  \tilde{H} &=&-\frac{1}{2} \frac{d^2}{dy_1^2} 
  + \frac{1}{2} \left.\frac{d^2V}{dy_1^2}\right|_{y_1=\tilde{y}} y_1^2
  + V(\tilde{y}).
\eeq
This effective Hamiltonian describes small $O(1)$ fluctuations 
about fixed point $\tilde{y}$.
The second derivative determines the excitation spectrum around the fixed 
point, and $ V(\tilde{y}) \sim O(\mu)$ is the energy of the bottom of the 
harmonic potential
well in which the system is localised.  In general, the potential will 
have more than one fixed point and an independent effective Hamiltonian
may be derived for each.
The way in which contributions from different fixed points 
combine to give the overall ground state of the quantum system will be 
treated for individual catastrophes.

The second collective mode $y_2$ is assumed to be simple harmonic, 
and thus the full Hamiltonian of the catastrophe model is
\beq
  H_\mathrm{cat} (\mathbf{y}) &=& 
  -\frac{1}{2} \frac{d^2}{dy_1^2} 
  -\frac{1}{2} \frac{d^2}{dy_2^2} 
  + V_\mathrm{cat}(y_1)  + \frac{1}{2} y_2^2,
  \label{Hcat}
\eeq
We relate the coordinates of the two
collective modes $\mathbf{y}$ to those of the bare modes 
$\mathbf{x}$ via the rotation
\begin{eqnarray}
  y_1 = c x_1 + s x_2,\quad y_2 = -s x_1 +c x_2,
\end{eqnarray}
where $c=\cos(\theta/2)$
and $s= \sin(\theta/2)$, and $\theta$ reflects the degree of mixing.   In 
terms of the ${\bf x}$-representation, $H_\mathrm{cat}(\mathbf{x})$ 
is not separable, 
and this 
rotation generates an interaction between the two bare modes 
${\bf x}$.  We quantise the collective coordinates $y_i$ and 
the bare coordinates $x_i$ according to 
\begin{eqnarray}
  y_i = 2^{-1/2}(b_i^\dag + b_i),\quad x_i = 2^{-1/2}(a_i^\dag + a_i),
\end{eqnarray}
with momenta defined canonically.  
In this second quantised notation, the two representations are  
related through a two-mode SU(2) squeezing transformation described by 
the unitary operator 
$W = \exp(-\frac{\theta}{2} a_1^\dag a_2 + \frac{\theta}{2} a_1 a_2^\dag)$.

To make the connexion with a familiar model: the above scheme is very similar to
the Dicke model in the thermodynamic limit.  Here, the two bare modes are 
the photon field and the collective atomic coordinate, and these 
are related to the collective excitations (polaritons) by just such a 
squeezing \cite{cetb03,note}.

In this paper, we consider two one-dimensional 
catastrophes --- the cuspoids $A_{+3}$ and $A_{+5}$, commonly referred to as 
the cusp and the butterfly.
We shall also consider a catastrophe that occurs in two dimensions, 
$V_\mathrm{cat}(y_1,y_2)$ and is non-separable.
In this case, we calculate the entanglement between the modes $y_1$ 
and $y_2$ with the catastrophe itself providing the interaction
between the modes.  In selecting which 
catastrophes to study, we require that the spectra of the catastrophe 
be bounded from below for all values of the control parameters at finite 
$\mu$.

\section{Entanglement about fixed points: 
  $\mu \to \infty$ limit \label{secinf}}

For the one-dimensional catastrophes, the two-mode 
Hamiltonian that determines the excitations 
about $\tilde{y}_1$ in the 
$\mu\to \infty$ limit is
\beq
 H = -\frac{1}{2}\frac{d^2}{dy_1^2} - \frac{1}{2}\frac{d^2}{dy_2^2} 
  +\frac{1}{2}\epsilon_1^2 y_1^2  + \frac{1}{2} y_2^2 
  +V(\tilde{y}_1)
  \label{eHam_wfn}
\eeq
with $\epsilon_1^2 = \left.d^2 V/dy_1^2\right|_{y_1=\tilde{y}}$.
The ground state wave function of the system is thus the Gaussian
\beq
  \Psi({\bf y}) = (\pi^2/\epsilon_1  )^{-1/4}
  \exp \rb{-\frac{\epsilon_1}{2}y_1^2 - \frac{1 }{2}y_2^2},
\eeq
which in the ${\bf x}$-representation reads
\beq
\Psi({\bf x}) =  \rb{\frac{\pi^2}{\epsilon_1}}^{1/4}
  \exp 
  \left\{
    -\frac{\epsilon_1}{2} (c x_1 + s x_2)^2
    -\frac{1 }{2} (s x_1 -c x_2)^2
  \right\}.
\eeq
To find the entanglement of this wave function, we require the reduced 
density matrix (RDM) of one of the bare modes, $x_1$, say.  This is obtained
through $\rho(x_1,x_1') = \int dx_2 \Psi(x_1, x_2) \Psi^*(x_1',x_2)$ as
\beq
  \rho(x_1,x_1') = \frac{\pi}
   {\sqrt{\epsilon_1  (s^2 \epsilon_1 +c^2 )}}
   \exp\left\{- \alpha (x_1^2 + {x_1'}^2) + \beta x_1 x_2\right\},
\eeq
where $\alpha$ and $\beta$ are coefficients, only the ratio of which 
is important for the entanglement:
\beq
   \frac{2\alpha}{\beta} = 
  \frac{(\epsilon_1 +1  )^2 + 2 \epsilon_1   
  \left[ \cot^2(\theta/2)+\tan^2(\theta/2) \right]}
  {(\epsilon_1 -  1)^2}.
  \label{ab}
\eeq
We shall quantify the entanglement in our two mode system with the von 
Neumann entropy $S$.  The entropy of the density matrix $\rho(x_1,x_1')$
is evaluated by comparison with the density matrix of a
harmonic oscillator at finite temperature.  Details of this approach have 
been given elsewhere \cite{NLCETB04}, and we just give the result here:
\beq
  S =  \frac{1}{\log 2}
  \left\{
    \frac{\Omega}{2T} \coth \rb{\frac{\Omega}{2T}}
    -\ln \left[2 \sinh  \rb{\frac{\Omega}{2T}} \right]
  \right\},
\eeq
where the ratio of frequency to temperature of the fictitious oscillator
is given by $\Omega/T = \mathrm{arccosh} (2 \alpha / \beta)$.
For the one-dimensional catastrophes, the entanglement is maximised 
when the squeezing angle is $\theta=\pi/2$.  
For this choice, Eq. (\ref{ab}) simplifies to
\beq
  \frac{2\alpha}{\beta} = 
  \frac{\epsilon_1^2 + 6 \epsilon_1 + 1}
  {(\epsilon_1 -  1)^2}.
  \label{ab2}
\eeq
This procedure is easily adapted to calculate the entanglement in 
the two-dimensional catastrophe.  

We now consider our three example catastrophes in turn.

\section{Cusp}
The cusp catastrophe, $A_{+3}$ is the most familiar and, 
from the point of view of 
applications, the most important catastrophe. 
With coefficients chosen for convenience, the scaled cusp 
potential is
\beq
  V_{+3}(y_1) =  \frac{1}{4\mu} y_1^4 + \frac{A}{2} y_1^2.
\eeq
We shall only consider a harmonic perturbation here, and reserve until 
later a discussion of the effects of linear perturbations.
We also shall set  $\theta=\pi/2$ here to give maximum 
mixing between the modes. This leaves us with a single control 
parameter $A$.

The full two-mode Hamiltonian in 
terms of the creation and annihilation operators of the 
$\mathbf{x}$-modes is
\beq
    H_{+3} ({\bf a}) &=&
  \frac{ A+3}
       {4}
  (a_1^\dag a_1 + a_2^\dag a_2 + 1)
  \nonumber\\
  &&+
   \frac{A-1}
       {8}
  ({a_1^\dag}^2 + a_1^2 + {a_2^\dag}^2 +  a_2^2)
  \nonumber\\
  &&+
  \frac{A-1}
       {4}
  (a_1^\dag a_2 + a_1 a_2^\dag+ a_1^\dag a_2^\dag + a_1 a_2 )
  \nonumber\\
  &&+
  \frac{1}{64 \mu}  (a_1^\dag + a_1 + a_2^\dag +  a_2)^4.
  \label{xcusp}
\eeq

It may at first appear unusual that the 
coefficient of $a_i^\dag a_i$ should depend
on the parameter $A$.
However,  it can be shown that, by individually squeezing the collective 
modes before applying the two-mode SU(2) transformation,
this dependence on $A$ can be removed. If both modes are 
squeezed identically, the entanglement properties of the system are
left invariant, since this squeezing then represents a global rescaling 
of the phase space.  For simplicity though, we retain the form of 
Eq. (\ref{xcusp}).

\begin{figure}[t]
\begin{center}
  \psfrag{S}{$S$}
  \psfrag{A}{$A$}
  \psfrag{Al0}{$A<0$}
  \psfrag{Ag0}{$A>0$}
  \psfrag{Ss}{$S^*$}
  \psfrag{As}{$A^*$}
  \psfrag{m}{$\mu$}
  \psfrag{m=10}{$\mu=10$}
  \psfrag{m=20}{$\mu=20$}
  \psfrag{m=40}{$\mu=40$}
  \psfrag{m=70}{$\mu=70$}  
  \psfrag{tdl}{$\mu\to \infty$}
  \includegraphics[width=1\linewidth,clip=true]{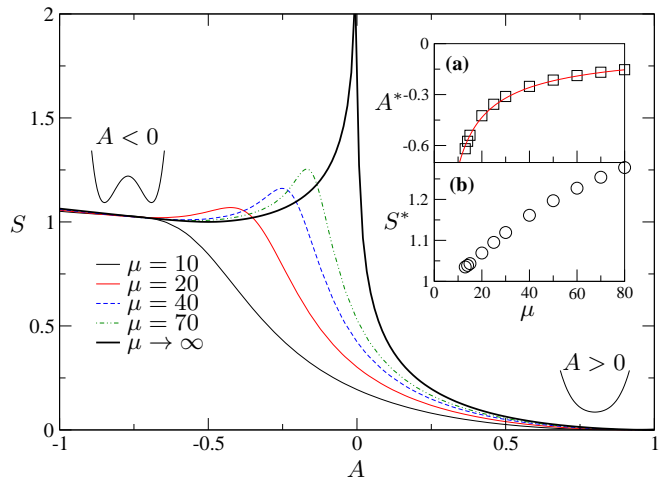}
  \caption{
    Entanglement properties of the cusp catastrophe.  
    The von Neumann entropy $S$ in the macroscopic limit 
    $\mu\to \infty$ (thick line)
    shows a divergence at the critical value $A=0$ where the potential 
    changes from a double- to a single-well structure (inset sketches).  
    Numerical results for finite $\mu$ show a peak near this point.  
    Inset {\bf (a)} shows the scaling with $\mu$ of the parameter value $A^*$
    at which the entanglement maximum occurs, and {\bf (b)} 
    shows the value of the entropy $S^*$ at this point.
    \label{figENT1}
  }
\end{center}
\end{figure}

We now consider the fixed points.
For $A > 0$, only one stable fixed point exists and this lies at the 
origin.  Taking $\mu \to \infty$, we see that the 
excitation energy about this fixed point is $\epsilon_1 = \sqrt{A}$.
For $A<0$, the origin becomes unstable, and two new stable fixed 
points appear at $y_1 =\pm \sqrt{\mu |A|}$. 
In the $\mu\to \infty$ limit, these two fixed points are 
degenerate and have the same excitation energy $\epsilon_1 = 2\sqrt{|A|}$.
The shape of the potential is sketched as insets in Fig. \ref{figENT1} 
and shows clearly the change of the potential from double to single 
well structure.
Note that the form $V_{+3}$, Eq. (11), is also used in Landau theory 
of phase transition in statistical mechanics, or in quantum  field 
theory ($\phi^4$-model). 
Describing the vanishing of the excitation energy as 
$\epsilon_1 \sim A^{z\nu}$, and the divergence of the ``correlation length'' 
as $\xi \equiv \epsilon^{-1/2} \sim A^\nu$, we find exponents 
$\nu = 1/4$ and $z=2$.  

We now consider the entanglement. 
For $A>0$, the entropy follows directly from the approach outlined in 
section \ref{secinf}.  For $A<0$, the situation is complicated slightly by 
the existence of two fixed points.  
With the limit $\mu\to\infty$ taken in correspondence 
with the thermodynamic limit, the ground state 
of the system would be an equal mixture of density matrices 
localised at the two 
fixed points.
We prefer here to use the limit $\mu\to \infty$
to calculate an approximate wave function for finite but large 
$\mu$.  This is obtained by taking a coherent superposition of 
the two localised wave functions and allows direct comparison 
with the numerical results for finite $\mu$. Since the two lobes 
are orthogonal, the reduced density matrix of the total system is equal 
to the sum of the reduced density matrices for the two lobes:  
$\rho_1 = 1/2 ( \rho_+ + \rho_-)$.
This is the same result as is obtained if one
takes the ground-state to be the incoherent mixture;
so the difference between these two approaches is unimportant. However, 
this will be seen  not to be the case when we consider the 
two-dimensional catastrophe.

From the general theory of entropy \cite{wehrl} we know that for 
$\rho = \sum_i \lambda_i \rho_i$ with $\lambda_i$ probabilities, the total
entropy $S(\rho)$ is bounded by
\beq
  \sum_i \lambda_i S(\rho_i) 
  \le  S(\rho) 
  \le  \sum_i \lambda_i S(\rho_i) - \lambda_i \lg \lambda_i
  \label{Sbound}.
\eeq
In the current situation, since $\rho_+$ is orthogonal to $\rho_-$, 
the upper bound becomes an equality.  Furthermore, since
$S(\rho_+) = S(\rho_-)$, 
we have $S(\rho_1) = S_{\rm mix} + S(\rho_+)$ with $ S_{\rm mix}=1$.
The mixing entropy represents the contribution from the 'global', i.e.
macroscopic, structure of the wave function, whereas local structure
enters through the individual $S(\rho_+)$ terms.
If the parity symmetry $V_{+3}(y_1)=V_{+3}(-y_1)$ is broken 
by an additional  linear term $\propto y_1$ in the potential, 
the degeneracy of the two fixed points would be lifted 
and the contribution from the mixing entropy $S_{\rm mix}=1$ would disappear.

The single-well entropy $S(\rho_+)$ is calculated as in section 
\ref{secinf}, and we plot the total entropy $S(\rho)$ in 
Fig. \ref{figENT1}.
The similarity between the behaviour of this simple cusp model and the 
QPT models is apparent.  At the critical point, the entropy diverges as
\beq
  S \sim \nu \lg A = \lg \xi,
\eeq
i.e., with the correlation length $\xi$,
and we thus see ``critical entanglement'' \cite{ON02}.

Numerically obtained results for finite $\mu$ are shown 
alongside the $\mu\to\infty$ result.  
The value of A for which the peak in the 
entanglement occurs at finite $\mu$, $A^*$,  
scales with $\mu$ to a very good 
approximation as $A^*= c \mu^{0.75}$ with a numerically 
determined constant of $c = 4.1$.  This relation is plotted in 
Fig. \ref{figENT1}a.  We mention that the exponent of $0.75 \approx 3/4$
has been observed numerically for the entropy in the Dicke model 
\cite{NLCETB03}.  We also investigated the value of the entropy $S^*$ 
at its peak (Fig. \ref{figENT1}b) but found no  convincing 
scaling relation for finite $\mu$.

\section{Butterfly}

The second one-dimensional catastrophe that we study is 
the butterfly, $A_{+5}$, which gives rise to the potential

\beq
  V_{+5}(y_1) = \frac{A_2}{2} y_1^2 + \frac{A_4}{4 \mu} y_1^4 
  + \frac{1}{6 \mu^2} y_1^6.
\eeq
The parameter space is two-dimensional ($A_2,A_4$), and rather than give a 
full account of this space, we simply look at two  representative values 
of $A_4$

{\it Case (i):} $A_4=0$.  For $A_2>0$, $y_1=0$ is the only fixed point 
and this has
excitation energy $\epsilon_1 = \sqrt{A_2}$.
For $A_2<0$, $\tilde{y} = \pm \sqrt{\mu} |A_2|^{1/4}$ are the two 
stable fixed points, both with $\epsilon_1 = \sqrt{2|A_2|}$.
Apart from numerical coefficients, 
the behaviour here is the same as that of the 
cusp.  This result generalises to all $A_{+k}$ catastrophes:
for $V_{+k}$ with $A_i=0, \forall i>2$ the excitation energy is 
$\sqrt{A_2}$ for $A_2>0$, and $ \sqrt{(k-3) |A_2|}$ for $A<0$, with 
behaviour like that of the cusp.

\begin{figure}[t]
\begin{center}
  \psfrag{S}{$S$}
  \psfrag{A2}{$A_2$}
  \psfrag{A2=0}{$A_2=0$}
  \psfrag{A2=1}{$A_2=1$}
  \psfrag{A2=2}{$A_2=2$}
  \psfrag{Ss}{$S^*$}
  \psfrag{As}{$A_2^*$}
  \psfrag{m}{$\mu$}
  \psfrag{m=5}{$\mu=5$}
  \psfrag{m=7}{$\mu=7$}
  \psfrag{m=10}{$\mu=10$}
  \psfrag{m=20}{$\mu=20$}  
  \psfrag{tdl}{$\mu\to \infty$}
  \includegraphics[width=1\linewidth,clip=true]{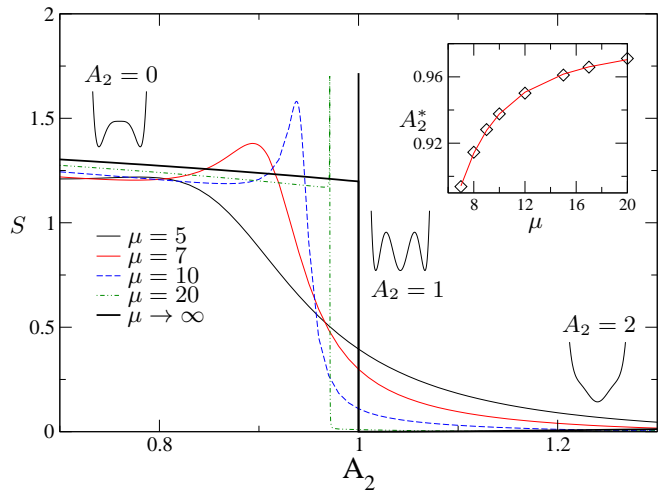}
  \caption{
    The von Neumann entropy of the Butterfly catastrophe with
    $A_4=-4/\sqrt{3}$ as the 
    potential undergoes a double-triple-single well transition, both for
    $\mu \to \infty$ and finite $\mu$.
    The profile of the entanglement is very different to that of the cusp 
    as the transition here is induced by a level crossing in the spectrum.  
    Inset shows scaling of $A_2^*$ as a function of $\mu$.
    \label{figENT2}
  }
\end{center}
\end{figure}

{\it Case (ii):} $A_4 = - 4 /\sqrt{3}$.  Here we see new behaviour absent 
in the cusp.  The $A_2$ parameter range is divided up
into three regions by the fixed points,
\beq
  A_2<0; &&  \tilde{y}=\pm \left[\frac{\mu}{\sqrt{3}}
        \rb{2+\sqrt{4 - 3 A_2}}
        \right]^{1/2} \equiv \tilde{y}_\pm
  \nonumber\\
  0<A_2<4/3; && \tilde{y}=0
    \nonumber\\
        && \tilde{y}= \tilde{y}_\pm
  \nonumber\\
  A_2 > 4/3; && \tilde{y}=0.
\eeq
Thus, increasing $A_2$ from below zero upwards, the potential moves 
through a sequence of first a double, then triple, then single well 
structures, as shown by the insets in Fig. \ref{figENT2}.

The stability or otherwise of  the  fixed points is only part of the 
story in determining the $\mu \to \infty$ ground state of the system.  
For $A_2>4/3$ and 
$A_2<0$, the situation is straightforward and the ground state 
is obtained exactly as for the two phases in the cusp.
In the central region $0<A_2<4/3$, however, we have three fixed points, 
and their weight in determining the ground state depends 
on the energy  $V(\tilde{y})$ of the bottom of the well at $\tilde{y}$.
In the
$\mu \to \infty$ limit, the system will be completely localised 
in whichever of the fixed points has the lowest base energy, or, if the 
energies are degenerate, we take an equal superposition 
to describe the large-$\mu$ wave function.  
For $A_2>1$,
$y=0$ is the fixed point with lowest energy, and for $A_2<1$ the 
two fixed points at finite displacements $y=\tilde{y}_\pm$ have the 
lowest energy and are degenerate.  Only at $A=1$ are all 
three points degenerate and we 
have a three-lobed wave function.  

This structure is induced by a level crossing in 
the $\mu\to \infty$ spectrum, with the energy of the double 
well crossing the energy of the single well at $A=1$.  
For finite $\mu$, the level-crossing is actually 
avoided, due to the overlap of all three wells.
This situation therefore bears some similarity to that
described in Ref. \cite{vid04}, where a discontinuous entanglement was 
observed at a level crossing associated with a first-order QPT.

Away from the level crossing, the entanglement is calculated just 
as for the cusp.  In the region of $A_2=1$, we need to exercise a 
little care, because the entanglement is discontinuous at $A_2=1$.
Exactly at this point, the 
excitation energies of the three wells do not disappear, but 
rather take the finite values $\epsilon_1 = (1,2,2)$.  
The entanglement in the central well (with $\epsilon_1=1$) is zero,
$S_0=0$, since the wave function is circularly symmetric about the origin 
($\epsilon_2=1$ as well) and can thus be written as a product state with 
respect to all co-ordinate systems.
The entanglement for each of the displaced wells
is $S_\pm \approx 0.197$.  Thus, by combining the appropriate 
density matrices, 
we find that for $A_2$ slightly less than unity, the double-well state has
$S=1.197$.  For $A_2$ just slightly bigger than unity we have $S=0$, 
due to the product state in the single well.  Directly at $A_2=1$ we have 
the three-lobed wave function, and 
$S = 2/3 S_+ + 1/2 S_- + \lg 3 \approx 1.716$.
These results plus the corresponding finite $\mu$ data are shown in
Fig. \ref{figENT2}.  The approach of the finite $\mu$ results to 
the $\mu\to\infty$ limit is nicely seen, and in particular 
to the limiting value of $S\approx 1.716$ at $A_2=1$.

We stress that the entanglement maximum occurs not at the value of $A_2$ 
at which the fixed point becomes unstable, but rather at the level 
crossing.  Moving through the points $A_2=0$ and $A_2=4/3$, where 
fixed point stability does change, nothing special happens to the 
entropy (or any other ground-state property), since these
fixed points do not contribute to the determination of the ground state
at these values of $A_2$.

By examining the finite $\mu$  data (Fig. \ref{figENT2}b), we determine that 
the value of $A_2$ at which the entanglement peak occurs scales 
as $A^*-1\sim c_0 \mu^{-c_1}$ with numerical 
parameters $(c_0,c_1)$ determined to be $(-3.55,1.90)$ to within 
a few percent.

\section{Two-dimensional Catastrophe}
\begin{figure}[t]
\begin{center}
  \psfrag{S}{$S$}
  \psfrag{g}{$\gamma$}
  \psfrag{gl1}{$\gamma<1$}
  \psfrag{gg1}{$\gamma>1$}
  \psfrag{gs}{$\gamma^*$}
  \psfrag{m}{$\mu$}
  \psfrag{m=10}{$\mu=10$}
  \psfrag{m=20}{$\mu=20$}
  \psfrag{m=30}{$\mu=30$}
  \psfrag{m=40}{$\mu=40$}  
  \psfrag{tdl}{$\mu\to \infty$}
  \includegraphics[width=1\linewidth,clip=true]
  {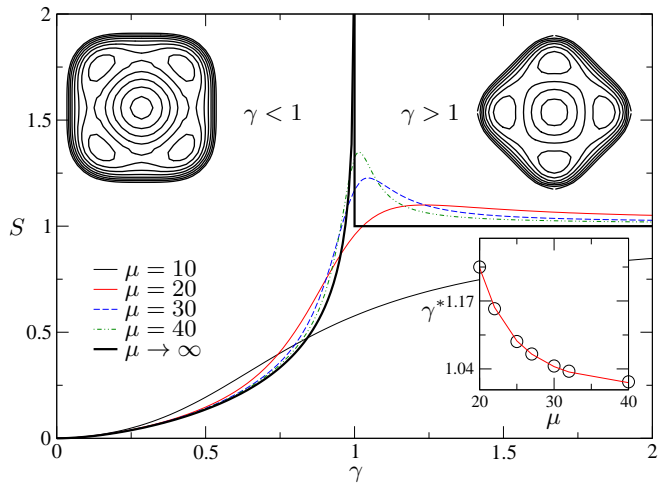}
  \caption{
    The von Neumann entropy of the two-dimensional molar 
    catastrophe with
    $A=-1$ as a function of $\gamma$.
    Plots of the potential for $\gamma<1$ and $\gamma>1$ are shown at 
    the top of the figure.  The origin of the potential is unstable
    and there are four stable potential wells satellite to this.
    Lower right inset shows scaling of $\gamma^*$ as a function of $\mu$.
    \label{figENT3}
  }
\end{center}
\end{figure}

The most familiar two-dimensional catastrophes are the umbillics 
with the germs $y_1^2 y_2 \pm y_2^3$.  
However, these are unsuitable for our purpose as their
spectra are not bounded from below and this, in fact, 
is true of all the two-dimensional,
elementary catastrophes of Thom \cite{thom}.
Therefore, we consider the non-simple catastrophe
\beq
 V_{\mathrm{m}} = \frac{1}{2} A(y_1^2 + y_2^2)
 + \frac{1}{4\mu}(y_1^4 +2 \gamma y_1^2 y_2^2 +y_2^4),
\eeq
where we have only included harmonic perturbations as before.  
This catastrophe
is described as non-simple because the germ (that part proportional to
$\mu^{-1}$ in the above) depends 
irreducibly on a modulus, $\gamma$, whereas 
simple germs have no free parameters.

The fixed point structure of $V_{\mathrm{m}}$ divides the behaviour into three
regimes in the $\mu\to \infty$ limit.  For  $A>0$, we obtain a
single fixed point at the origin,  and since the ground-state 
of the system is a product state of two Gaussians with the same width,
there is no entanglement.
For $A<0$, the origin is unstable; for $\gamma \ne 1$, 
the system possesses four fixed points,
as is readily observed from the molar-shaped potentials plotted as
insets  of Fig. \ref{figENT3}.  For all $\gamma>1$, 
the four stable fixed points 
lie on the lines $y_1=0$ and $y_2=0$, whereas for $\gamma<1$ they lie 
on the diagonals $y_1=\pm y_2$.  In the following,
we set $A_2=-1$ throughout, as the entanglement 
properties are the same for all $A_2<0$.  We calculate 
the entanglement between modes $y_1$ and $y_2$ induced by the 
interaction in the catastrophe itself, and do not apply the two-mode 
squeezing.

We first study $\gamma >1$ as this is
the simpler of the two cases.  The stable 
fixed points are given by
\beq
  (y_1, y_2) = (\pm \sqrt{\mu},0)
  ;\quad
  (y_1, y_2) = (0,\pm \sqrt{\mu}).
\eeq
At each fixed point, $y_1$ and $y_2$ are the excitation
coordinates with excitation energies
\beq
  \epsilon_+^2 = 2;\quad  \epsilon_-^2 =  \gamma-1.
\eeq
Excitations in the direction of the displacement $\pm \sqrt{\mu}$
are described $\epsilon_+$.

The individual wave functions localised around any of these fixed points 
are unentangled, since they are just products of Gaussians is the $y_1$
and $y_2$ directions.  However, combining these four functions into the 
four-lobed wave function that describes the large $\mu$ limit,
the total system is entangled.
This is solely due to the mixing entropy of its four lobed structure.

We can not calculate the entanglement of this structure in the way we did 
for the one-dimensional catastrophes, because the four reduced density 
matrices of each lobe are not orthogonal.  This means that 
the upper bound in Eq. (\ref{Sbound}) remains as an upper bound, 
and is not equality.  
Nevertheless, we can proceed as follows. 
Writing $\ket{\tilde{y}_1,\tilde{y}_2}$ for the wave function 
of the system localised at $(\tilde{y}_1,\tilde{y}_2)$,
the four-lobed large-$\mu$ wave function can be written as
\beq
  \ket{\Psi} &= &
  \frac{1}{2}
  \left\{ 
    \ket{\tilde{y}, 0}
    +\ket{-\tilde{y},0} 
    +\ket{0,\tilde{y}}
    +\ket{0,-\tilde{y}}
  \right\}
\eeq
with $\tilde{y}=\sqrt{\mu}$.
Given that the individual lobes contribute nothing to the 
entanglement by themselves,  we ignore their individual structure 
in this description.
In the limit $\mu\to \infty$, the three single-mode states
$\ket{0},\ket{\pm\tilde{y}}$ are all orthogonal, and thus the RDM of 
one of the modes 
$\rho_1 = \mathrm{Tr}_2\ket{\Psi}\bra{\Psi}$ 
is
\beq
  \rho_1 = \frac{1}{4} 
  \left\{
    \rb{\frac{}{}\ket{\tilde{y}}+\ket{-\tilde{y}}}
    \rb{\frac{}{}\bra{\tilde{y}}+\bra{-\tilde{y}}} 
    + 2 \ket{0}\bra{0}
  \right\}.
\eeq
Furthermore, the orthogonality of these states means that this 
density matrix can be simply treated as a three-by-three matrix
and the 
entropy is simply $S=1$, independent of $\gamma$ for $\gamma>1$. 

It is interesting to note that had we taken as the ground-state 
density matrix the incoherent mixture of the four 
contributions, 
\beq
  \rho &=& \frac{1}{4}
  \left\{
    \op{\tilde{y},0}{\tilde{y},0} +  \op{-\tilde{y},0}{-\tilde{y},0} 
  \right.
  \nonumber\\
  &&~~~~~~~~~~~~
  \left.
    +\op{0,\tilde{y}}{0,\tilde{y}} +  \op{0,-\tilde{y}}{0,-\tilde{y}} 
  \right\},
\eeq
leading to the RDM
\beq
  \rho_1 =\frac{1}{4}
  \left\{
  \op{\tilde{y}}{\tilde{y}} +   \op{-\tilde{y}}{-\tilde{y}} + 2\op{0}{0}
  \right\}
\eeq
and a value of the von Neumann entropy of $S=3/2$, which 
is clearly at variance with the numerical results.

We now consider the region $\gamma <1$, and for simplicity we also assume 
$\gamma>0$.  The four fixed points are
\beq
  (y_1,y_2) = \rb{\pm\sqrt{\frac{\mu}{1+\gamma}}, 
                  \pm\sqrt{\frac{\mu}{1+\gamma}}}
\eeq
where the two $\pm$ signs are independent.  Each fixed point has 
the excitation energies
\beq
  \epsilon_+^2 = 2
  ;\quad
  \epsilon_-^2 = 2\frac{1-\gamma}{1+\gamma}.
\eeq
The eigenmodes of the system are not $y_1$ and $y_2$, but rather lie 
along, and perpendicular to, the diagonals of the $y_1$-$y_2$ plane. 
Each individual fixed-point wave function is thus entangled with 
respect to modes $y_1$ and $y_2$.

This entanglement can be calculated as in section \ref{secinf}, but here 
with two excitation energies and the rotation between the 
eigenmodes and the ${\bf y}$ coordinates.  The entanglement determining 
parameter 
$2\alpha/\beta$ is evaluated to be
\beq
  \frac{2\alpha}{\beta} = \frac{4 - 3 \gamma^2 + 4 \sqrt{1-\gamma^2}}
  {\gamma^2},
\eeq
from which the single-lobe entanglement  follows directly.

The contribution of the four-lobed structure of the large-$\mu$ 
superposition can be assessed as follows.
From a macroscopic point of view, we can ignore the structure of 
the individual lobes, and write the wave function as
\beq
  \ket{\Psi} &=& \frac{1}{2} 
  \left\{
    \ket{\tilde{y},\tilde{y}}   + \ket{\tilde{y}-,\tilde{y}}
    +  \ket{-\tilde{y},\tilde{y}} +   \ket{-\tilde{y},-\tilde{y}}
  \right\}
  \nonumber\\
  &=& \rb{\frac{}{} \ket{\tilde{y}} + \ket{-\tilde{y}}}\otimes
   \rb{\frac{}{} \ket{\tilde{y}} + \ket{-\tilde{y}}}.
\eeq
The second forms clearly shows this wave function to be a product
state from the macroscopic viewpoint.  Thus the mixing entropy of 
forming the four-lobed structure is zero, and the entropy of the 
system is just the single lobe entropy above.

In Fig. \ref{figENT3} we plot these results alongside 
the numerical data for finite $\mu$.
The scaling of $\gamma^*$ with $\mu$ is observed to be 
$\gamma^*-1 = c_0 \mu^{-c_1}$ with coefficients fitted as
$(c_0,c_1) = (4.93\times10 ^{4},4.09)$.


\section{Conclusions}

We have constructed and studied a family of quantum catastrophe 
models, and investigated their ground-state entanglement properties.
The cusp catastrophe, with its bifurcating fixed point, demonstrates 
behaviour that is remarkable similar to the QPT models, such as the Dicke 
model --- underlining the importance of bifurcations of classical 
fixed points in this context.  It should be noted that whilst this 
bifurcation occurs for all values of $\mu$, a peak in the entanglement
is only observed when $\mu$ is sufficiently large ($\mu >10$ here).  This
illustrates that the bifurcation is not, in itself, a sufficient condition
for the occurrence of the entanglement maximum, 
but that the system must also be capable of sufficient delocalisation. 
The butterfly catastrophe displays very different behaviour to the cusp ---
namely a discontinuous entropy induced by a level crossing 
in the macroscopic limit.

The cusp and the two-dimensional catastrophe demonstrate that a 
mixing term in the entropy can contribute to the total entanglement 
in cases where a wave function is split up into  localisation areas 
that are separated within (abstract) position space. In 
particular the two-dimensional catastrophe suggests a 
distinction between `global' and `local' (within the lobes) 
entanglement, and one could speculate that in more complex 
situations, with wave functions split up further and
further, a hierarchy of entanglement entropies might emerge.

Our  results also have a bearing on the issue of quantum chaos
and entanglement in such systems, as the model here is capable of
emulating the behaviour of  more sophisticated nonlinear Hamiltonians,
despite being separable --- and thus integrable.  
It is clear that there is
no unequivocal relation between delocalization  and the onset
of quantum chaos on one hand and the peaking of entanglement on the other.

This work was supported by the Dutch Science Foundation NWO/FOM 
and the UK EPSRC Network `Transport, Dissipation, and Control 
in Quantum Devices'.



\begin{thebibliography}{99}
\bibitem{note} A degree of single-mode squeezing is also required 
  in the superradiant phase of the Dicke model.
\end{thebibliography}
\end{document}